\documentclass[aip,graphicx]{revtex4-2}

\usepackage{graphicx}
\usepackage{amsmath}
\usepackage{booktabs}

\begin{document}

\title{Type-1.5 SNSPD: Interacting vortex theory of two bandgap superconducting single photon detectors} 

\author{Leif Bauer}\altaffiliation{These authors contributed equally to the work}
\author{Daien He}\altaffiliation{These authors contributed equally to the work}
\author{Sathwik Bharadwaj}
\affiliation{The Elmore Family School of Electrical and Computer Engineering, Purdue University, West Lafayette, 47907, IN, USA}
\author{Shunshun Liu}\affiliation{Department of Materials Science and Engineering, University of Virginia, Charlottesville, VA, 22903, USA}
\author{Prasanna V. Balachandran}
\affiliation{Department of Materials Science and Engineering, University of Virginia, Charlottesville, VA, 22903, USA}
\author{Zubin Jacob}\email{zjacob@purdue.edu}
\affiliation{The Elmore Family School of Electrical and Computer Engineering, Purdue University, West Lafayette, 47907, IN, USA}

\date{\today}

\begin{abstract}
Photon detectors based on type-2 superconductors have found widespread applications from on-chip quantum computing to quantum remote sensing. Here, we develop the theory for a new class of type-1.5 superconducting nanowire single photon detectors (SNSPDs) based on two bandgap superconductors with high transition temperatures such as MgB$_2$ ($T_c$ $\sim$38.6$K$). We show that vortex-vortex interactions in two component condensates lead to a unique operating regime where single photons can seed multiple vortices within a hotspot. We also show that dark counts are suppressed in the type-1.5 regime compared to the widely studied type-2 SNSPDs. Our work opens the door for exploring the unique vortex physics of two-gap superconductors for quantum device applications.
\end{abstract}


\maketitle 
Superconducting nanowire single photon detectors (SNSPDs) have found success in numerous applications including on-chip quantum computing \cite{alexander_manufacturable_2025,wang_boson_2019,kaur_-chip_2025}, quantum remote sensing \cite{bao_photon_2024,blakey_quantum_2022}, and on-chip spectroscopy \cite{cheng_broadband_2019}. Generally,  SNSPDs have used type-2 superconducting materials such as NbN, WSi, or MoSi. These devices have demonstrated state-of-the-art performance in sensitivity \cite{reddy_superconducting_2020} and timing resolution \cite{korzh_demonstration_2020} across a wide range of visible and infrared wavelengths \cite{verma_single-photon_2021,marsili_efficient_2012}. However, improvements in photon detection are still necessary to increase operating temperature and increase detection wavelength. One approach to realizing the next generation of SNSPDs is to exploit novel superconducting materials which may provide an avenue for high temperature operation through unique single photon detection mechanisms. 

Photon detection in SNSPDs begins with a reduction in the superconducting order parameter due to the photon-induced hotspot \cite{natarajan_superconducting_2012,kozorezov_quasiparticle-phonon_2000}. For specific device geometries, this causes a vortex to cross the width of the nanowire which disturbs the local phase leading to destruction of the superconducting state \cite{engel_detection_2015,jahani_probabilistic_2020,embon_imaging_2017,akhlaghi_reduced_2012}. The magnitude of bias current significantly affects the probability of detection. Therefore, SNSPDs are often biased close to the critical current to improve detection efficiency\cite{esmaeil_zadeh_single-photon_2017}. However, latent thermal energy can also cause vortex crossing events to occur  \cite{jahani_probabilistic_2020}. These events, also called dark counts, become particularly prevalent at high bias currents where the probability of vortex crossing is increased \cite{bulaevskii_vortex-induced_2011}. The combined effects of dark count rate and detection efficiency determine the minimum detectable power\cite{hadfield_single-photon_2009}. Therefore, the reduction of dark counts at high bias currents can improve sensitivity, and likewise increase operating temperature.

Superconducting materials with unique vortex physics are interesting candidates for the next generation of SNSPDs. Recently, vortices in MgB$_2$ were discovered to have both long-range attraction and short-range repulsion \cite{moshchalkov_type-15_2009}. This behavior has been called type-1.5 superconductivity, and occurs due to the presence of two superconducting bandgaps ($\pi$-band and $\sigma$-band). The presence of two bandgaps leads to two separate order parameters $\psi_1$ and $\psi_2$. In clean MgB$_2$, the $\pi$-band operates in the type-1 regime and the $\sigma$-band operates in the type-2 regime. Due to this combination, the total order parameter has properties of both type-1 and type-2 materials, causing both attractive and repulsive vortex-vortex interactions to occur. This opens up a unique and intriguing question of whether type-1.5 superconductors can be exploited for SNSPDs.

In this paper, we develop an ab-initio theory of multiband SNSPDs operating in this unique type-1.5 regime. We demonstrate that type-1.5 SNSPDs display unique properties such as single photon induced nucleation of two-vortex clusters, and a reduced barrier for two-vortex crossing. We also find that clean MgB$_2$ operating in the type-1.5 regime has significantly suppressed dark counts compared to type-2 MgB$_2$, resulting in improved sensitivity. 

Our focus in this paper is on MgB$_2$, however our model is applicable to other type-1.5 superconductors. We note that MgB$_2$ has several unique material properties of interest for device applications. It has the highest critical temperature of BCS superconductors at 38.6$K$ \cite{moshchalkov_type-15_2009}, and the smallest magnetic penetration depth ($\lambda = 56.8nm$) demonstrated in thin film superconductors. The small magnetic penetration depth is a result of MgB$_2$’s uniquely small normal state resistivity which is in part explained by its large electron diffusion \cite{tarantini_effects_2006}. Recently, SNSPDs fabricated from MgB$_2$ have demonstrated improvements in several device metrics, such as reset times as small as 130ps \cite{cherednichenko_low_2021} and photon response at bias temperatures up to 20$K$ \cite{charaev_single-photon_2024}.

We will first briefly compare normal state formation of type-1 and type-2 superconductors. Type-1 superconductors exhibit a first order phase transition with magnetic field while type-2 superconductors exhibit a second order phase transition. In the type-1 intermediate state, where normal and superconducting states both persist, the energy per unit area of superconducting-normal interface is positive \cite{tinkham_introduction_2015}. This leads to normal cores combining to minimize the interface area. In type-2 materials, the interface energy is negative, leading to a splitting of normal regions into a lattice of normal cores each with a single magnetic flux quantum (i.e. vortices). However, there are some similarities between the type-1 macroscopic normal domains and the type-2 vortices. In the type-1 intermediate state, macroscopic normal domains contain quantized flux as demonstrated in the Little-Parks experiment \cite{little_observation_1962}. This captured flux causes circulations of current around the macroscopic domains similar to vortices. Additionally, interactions between the type-1 quantized flux can be treated as attractive \cite{kramer_thermodynamic_1971}. This explains the behavior of vortices in type-1.5 superconductors, where vortices experience both long-range attraction and short-range repulsion due to the combination of type-1 and type-2 order parameters.

There has been some debate over type-1.5 superconductors and two component Ginzburg-Landau theory due to the inclusion of multiple coherence lengths \cite{kogan_ginzburg-landau_2011,babaev_comment_2012}. Several theoretical studies have demonstrated that Ginzburg-Landau models reduce to a single coherence length in the limit as $T\rightarrow T_c$\cite{kogan_ginzburg-landau_2011,silaev_microscopic_2012}. However, there have been subsequent microscopic studies based on Eilenberger \cite{silaev_microscopic_2012,silaev_microscopic_2011} and Boguliubov de Gennes \cite{timoshuk_microscopic_2024} models demonstrating that type-1.5 behavior does occur at all other temperatures $0<T<T_c$\cite{silaev_microscopic_2011} for small interband couplings $\lambda_{12}<0.1$\cite{silaev_microscopic_2012}. Additionally, Usadel theory has been used to show that two band superconductors with impurities can display type-1.5 behavior for large interband couplings \cite{garaud_properties_2018}. Meanwhile, several experiments have demonstrated vortex clustering at low-temperatures ($T<0.5T_c$) in single crystal MgB$_2$ \cite{gutierrez_scanning_2012,nishio_scanning_2010,moshchalkov_type-15_2009} and single crystal Sr$_2$RuO$_4$ \cite{bjornsson_scanning_2005}. These experiments utilize a variety of methods including SQUID on tip \cite{nishio_scanning_2010}, Bitter decoration \cite{moshchalkov_type-15_2009}, and Hall probe microscopy \cite{gutierrez_scanning_2012}. Therefore, we will assume in the following discussions that a two component Ginzburg-Landau model is applicable in MgB$_2$ at temperatures below $0.5T_c$. Although this paper focuses on MgB$_2$, the model we present is generalizable to other type-1.5 superconductors. 

\begin{figure}[ht!]
\centering
\includegraphics[width=\textwidth]{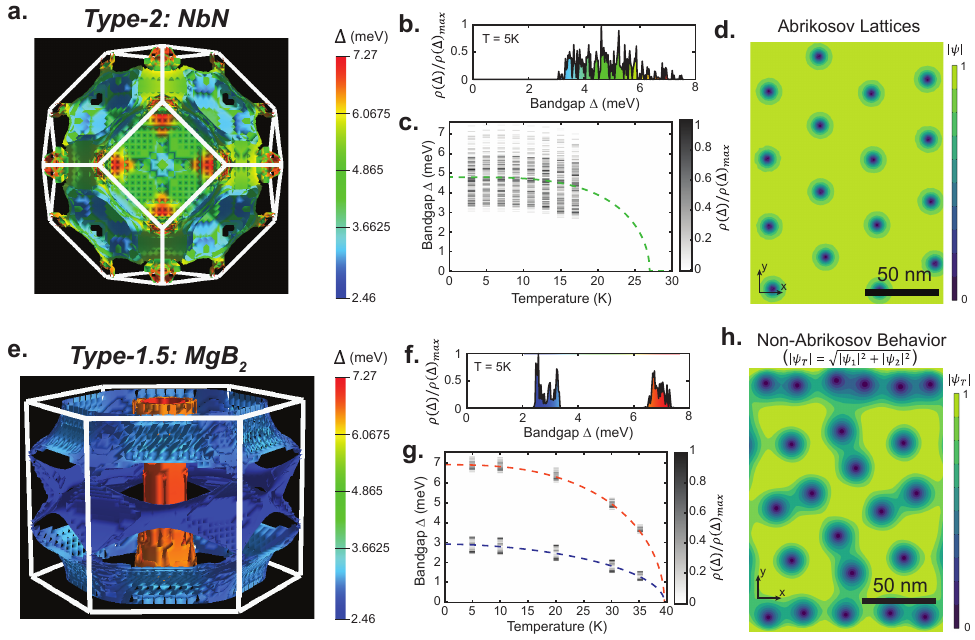}
\caption{Comparison of type-2 (NbN) and type-1.5 (MgB$_2$) superconductors. (a) Projection of NbN superconducting bandgap at $T = 5K$ onto Fermi surface. (b) Normalized distribution of NbN superconducting gap. (c) Temperature dependence of NbN superconducting bandgap distribution demonstrating a single gap. (d) The order parameter in NbN, a type-2 superconductor, displays an Abrikosov lattice structure under the presence of a magnetic field.  (e) Projection of MgB$_2$ superconducting bandgap at $T = 5K$ onto Fermi surface showing two distinct surfaces. The orange surface maps to the $\sigma$-band and the blue surface maps to the $\pi$-band . (f) Normalized distribution of MgB$_2$ superconducting gap. (g) Temperature dependence of MgB$_2$ superconducting bandgap distribution demonstrating two distinct bandgaps. (h) In MgB$_2$ under the presence of a magnetic field, the order parameter displays complex non-Abrikosov behavior due to the competition of attractive and repulsive forces on the vortices.}
\label{Fig:Type1_5vType2}
\vspace{-10pt}
\end{figure}

We begin our comprehensive interacting vortex model with density functional theory (DFT) calculations of the superconducting bandgap and Eliashberg electron-phonon coupling parameters. We utilize these parameters as well as those found from experiment in our time-dependent Ginzburg Landau (TDGL) simulations. These simulations capture the behavior of vortices under an applied magnetic field, current, or hotspot. For the MgB$_2$ TDGL simulations we use a two band model which has two order parameters $\psi_1$ and $\psi_2$ for the $\sigma$-band and $\pi$-band  respectively. Cooper pairs from the two bands are coupled via a Josephson-type interaction with a fixed phase difference of either $0$ or $\pi$\cite{geurts_vortex_2010}. In Fig.\ \ref{Fig:Type1_5vType2}, we compare a type-2 superconductor (NbN) to a type-1.5 superconductor (MgB$_2$). Comparing Fig.\ \ref{Fig:Type1_5vType2}a,b,c to Fig.\ \ref{Fig:Type1_5vType2}e,f,g (calculated from DFT), we see that the superconducting bandgap distribution of NbN falls along a single region, while the superconducting bandgap distribution of MgB$_2$ has two distinct regions (orange and blue) representing the $\sigma$ and $\pi$ gaps respectively. Both materials follow BCS theory as demonstrated by the temperature dependent gap in Fig.\ \ref{Fig:Type1_5vType2}c,g \cite{tinkham_introduction_2015,choi_origin_2002}. 

We compare the TDGL results for the order parameter under an applied magnetic field in Fig. \ref{Fig:Type1_5vType2}d,h. In Fig.\ \ref{Fig:Type1_5vType2}d we calculate the order parameter from TDGL for NbN under an applied magnetic field of $H=0.46H_{c,2}$ \cite{alstrom_magnetic_2011}. The vortices in NbN form an Abrikosov lattice due to the repulsive vortex-vortex interactions present in type-2 materials \cite{kramer_thermodynamic_1971}. In Fig.\ \ref{Fig:Type1_5vType2}h we plot the combined order parameter $|\psi|=\sqrt{|\psi_1|^2 +|\psi_2|^2}$ from the two band TDGL of MgB$_2$ under an applied magnetic field of $H=0.78H_{c,2}$ \cite{chan_analysis_2007}. Here we see vortices cluster due to the combination of short-range repulsion and long-range attraction which leads to non-Abrikosov behavior \cite{moshchalkov_type-15_2009}.

The multiband nature of MgB$_2$ as well as the type-1.5 regime leads to device behavior beyond what can be represented with the London model. The dark count rate $D$ is related to the maximum energy barrier for vortex crossing $U_{max}$ by the following equation\cite{engel_detection_2015,jahani_probabilistic_2020}
\begin{equation}
  D = \alpha e^{-U_{max}/k_BT}
  \label{Eq:DCR}
\end{equation}
where $\alpha$ is the vortex attempt rate, $k_B$ is Boltzmann's constant, and $T$ is the temperature. In simple scenarios this energy barrier can be calculated from London theory using the following equation\cite{kogan_interaction_2007,bulaevskii_vortex-induced_2011}
\begin{equation}
    U_{max}/\varepsilon_0 = \max_{x_\nu}\left[\ln\left(\frac{2W}{\pi\xi}\right)\sin(\pi x_\nu)-\frac{I}{I_c}\frac{2Wx_\nu}{\exp(1)*\xi}\right]
    \label{Eq:London}
\end{equation}
where $\epsilon_0$ is the vortex energy, $W$ is the nanowire width, $\xi$ is the Ginzburg-Landau coherence length, $x_\nu$ is the vortex position with range of $[0,1]$ over x positions of $[0,W]$, and $I/I_c$ is the bias current normalized by the critical current. 

London theory contains a single magnetic penetration depth and therefore is not applicable for multiband superconductors or type-1.5 superconductors. Therefore, we use TDGL simulations to calculate the vortex barrier while capturing the complicated nature of multiband vortex-vortex interactions. Utilizing the string method \cite{e_string_2002}, the free energy from vortices placed at saddle points (i.e.\ stationary positions) represents the maximum potential barrier faced in a vortex crossing \cite{vodolazov_saddle_2012}. Therefore, $U_{max}$ can also be calculated from TDGL using the following equations \cite{vodolazov_saddle_2012,qiu_numerical_2008,chan_analysis_2007}
\begin{equation}
    U_{max} = F_{saddle}-F_{ground}-\frac{\hbar}{2e}\frac{I}{I_c}\Delta\varphi
\end{equation}
\begin{equation}
    F(\psi) = \int(F_1+F_m)d^3x
    \label{Eq:1band_Free_energy}
\end{equation}
\begin{equation}
    F(\psi_\sigma,\psi_\pi) = \int(F_\sigma+F_\pi+F_{\sigma\pi}+F_m)d^3x
    \label{Eq:2band_Free_energy}
\end{equation}
where $F_{saddle}$ is the free energy at the saddle point (Eq.\ \ref{Eq:1band_Free_energy} for NbN and Eq.\ \ref{Eq:2band_Free_energy} for MgB$_2$), $F_{ground}$ is the free energy with no vortices, and $\Delta\varphi$ is the change in phase from the ground state across the nanowire length. $F_1$, $F_\sigma$, and $F_\pi$ are the free energies from the order parameters in their respective bands, $F_{\sigma\pi}$ is the energy from interband Josephson coupling, and $F_m$ is the free energy in the magnetic field (see Supplementary Materials).

\begin{figure}[ht!]
\centering
\includegraphics[width=\textwidth]{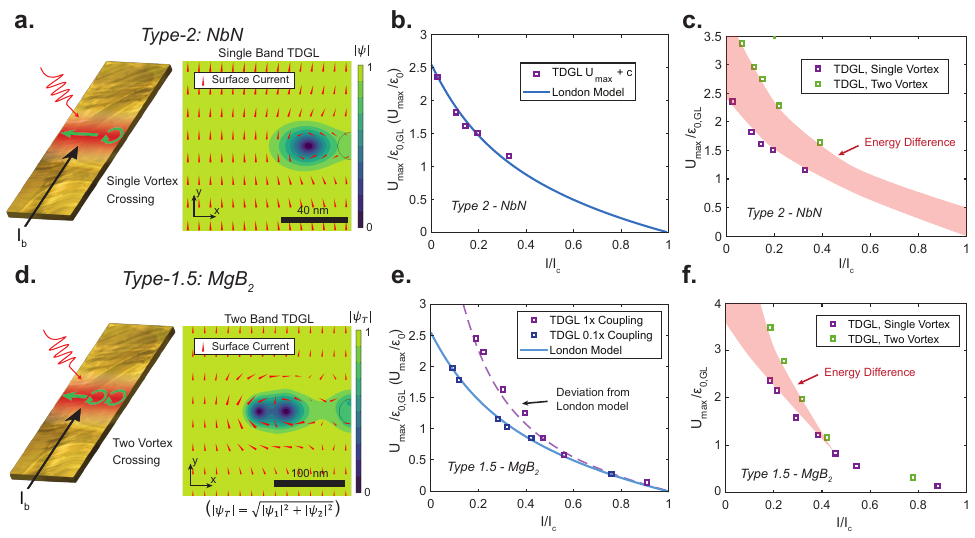}
\caption{Comparison of type-2 and type-1.5 SNSPD behavior. (a) Schematic and TDGL simulation of NbN with hotspot formed at edge of film. The hotspot nucleates a vortex which crosses the film, causing the superconductor to transition to the normal state. (b) Normalized single-vortex barrier in NbN with added constant ($c = 0.6$) calculated from TDGL matching closely with normalized London model ($U_{max}/\varepsilon_0$). (c) Comparison of single-vortex and two-vortex normalized vortex barrier in NbN with added constant, demonstrating a larger energy for two-vortex crossing at high bias currents. (d) Schematic and TDGL simulation of MgB$_2$ with hotspot formed at edge of film. The hotspot nucleates a two-vortex cluster which crosses the film, causing the superconductor to transition to the normal state. (e) Normalized single-vortex barrier in MgB$_2$ calculated from two-band TDGL deviating from normalized London model ($U_{max}/\varepsilon_0$). When the interband coupling $\eta$ is reduced by a factor of 10, the TDGL single-vortex barrier matches closely with the London model. (f) Comparison of single-vortex and two-vortex normalized vortex barrier in MgB$_2$, demonstrating similar energy for two-vortex crossing at high bias currents.}
\label{Fig:BarrierDifferences}
\end{figure}

In Fig.\ \ref{Fig:BarrierDifferences} we compare the vortex crossing behavior of type-2 and type-1.5 SNSPDs calculated via TDGL and from London theory. Vortex crossing can be directly simulated in TDGL by nucleating a vortex via a diffusive hotspot \cite{zotova_photon_2012}. We use TDGL in Fig.\ \ref{Fig:BarrierDifferences}a,d  to calculate the SNSPD response to a diffusive hotspot formed at the edge of the nanowire under a bias current $I_b$. In Fig.\ \ref{Fig:BarrierDifferences}a, the diffusive hotspot nucleates a single-vortex in NbN, leading to a vortex crossing event, which then breaks the superconducting state. However, in type-1.5 MgB$_2$ we find that the hotspot nucleates a two-vortex cluster as shown in Fig.\ \ref{Fig:BarrierDifferences}d, which then crosses the nanowire and breaks the superconductor. Additionally, we note that although single band TDGL vortex barriers matches closely with the London model, two-band TDGL vortex barriers deviate from the London model at low bias currents (see Fig.\ \ref{Fig:BarrierDifferences}b,e). When the interband coupling $\eta$ is reduced significantly, the two-band TDGL vortex barrier recovers the behavior predicted by London theory  (see blue squares in Fig.\ \ref{Fig:BarrierDifferences}e). Lastly, in Fig.\ \ref{Fig:BarrierDifferences}c,f we find that although an energy difference persists between single and two-vortex barriers in type-2 systems, the energy difference disappears in type-1.5 systems at larger bias currents ($I/I_c>0.4$). This indicates that two-vortex events may also contribute to the dark counts and photon counts in type-1.5 SNSPDs.

\begin{figure}[htb!]
\vspace{-30pt}
\centering
\includegraphics[width=\textwidth]{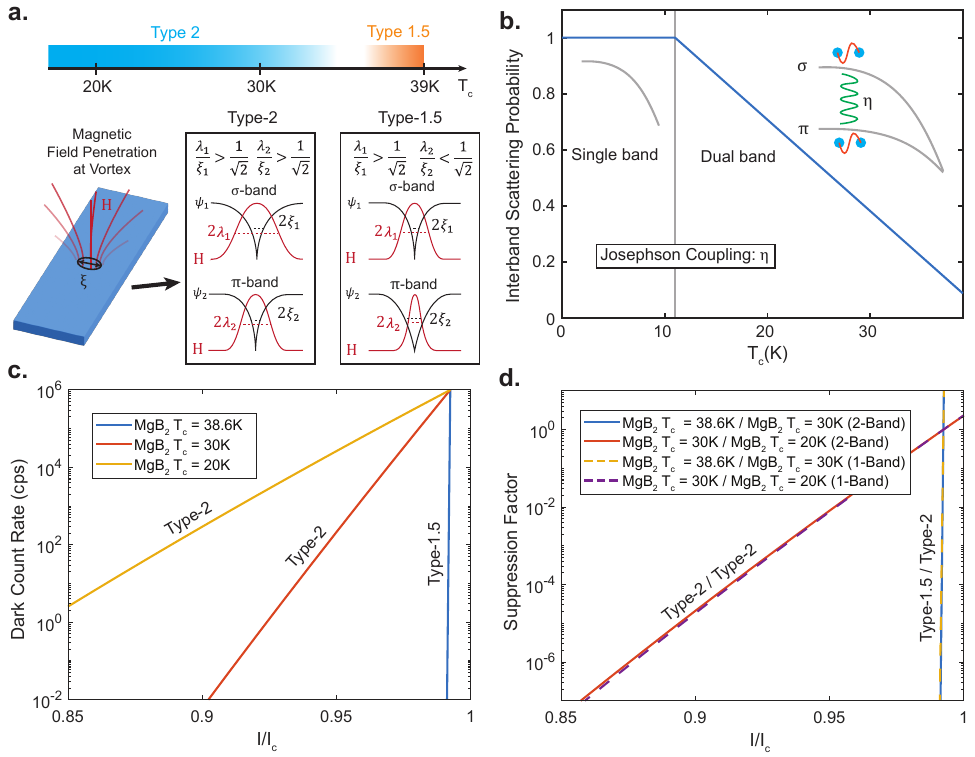}
\caption{Dark count suppression in type-1.5 SNSPDs. (a) Comparison of type-2 and type-1.5 order parameters and magnetic field penetration in two-band MgB$_2$ SNSPDs. Dashed lines represent length scales of coherence length and magnetic penetration depth. Clean MgB$_2$ close to 38.6$K$ displays type-1.5 behavior due to a type-1 $\pi$-band and type-2 $\sigma$-band . As $T_c$ of MgB$_2$ is reduced, the normal state resistivity increases, leading to an increase in penetration depth in the $\sigma$-band and $\pi$-band. This leads to a transition to type-2 behavior in both bands at lower $T_c$ ($T_c$$\lesssim$35K). (b) Change in interband scattering probability $a$ with $T_c$. An increase in interband scattering probability occurs as $T_c$ is reduced, which then approaches 1 at $T_c=11K$. (c) Comparison of the dark count rates of type-1.5 SNSPDs to type-2 SNSPDs. Type-1.5 SNSPDs show significantly sharper current dependence, indicating lower dark count rates at high bias currents. (d) Suppression factor of type-1.5 SNSPD is compared to type-2 SNSPDs. Type-1.5 SNSPD displays significantly more suppression compared to type-2 SNSPDs, even with large difference in $T_c$. This suppression remains in the single band case shown in dashed lines ($\eta = 0, a = 1$).}
\label{Fig:DCR_Differences}
\vspace{-10pt}
\end{figure}

Type-1.5 behavior with short-range repulsion and long-range attraction has so far only been demonstrated in clean MgB$_2$ with high critical temperatures near 38.6$K$ \cite{moshchalkov_type-15_2009}. However, MgB$_2$ can also display type-2 behavior in dirtier samples with reduced $T_c$ \cite{curran_spontaneous_2015}. As $T_c$ decreases the normal state resistivity increases due to an increase in interband and intraband scattering \cite{putti_intraband_2007}. This leads to an increase in the magnetic penetration depth of the $\pi$-band, changing the type-1 $\pi$-band to type-2. Thus, MgB$_2$ SNSPDs with significantly reduced $T_c$ (i.e.\ $T_c$ $\lesssim$ 35K) switch from type-1.5 vortices to type-2 vortices (see Supplementary Materials). We demonstrate these changes in MgB$_2$'s two component vortices in Fig.\ \ref{Fig:DCR_Differences}a.

The two-band nature of MgB$_2$ also changes with critical temperature. In theory\cite{putti_intraband_2007} and experiment\cite{putti_observation_2006,di_capua_role_2007} it has been demonstrated that MgB$_2$ remains a two-band system until $T_c=11K$. We take the change in superconducting bandgaps to cause a linear change in the $\sigma\rightarrow\pi$ interband scattering probability $a$. When the bands combine the scattering probability approaches 1 (see Fig.\ \ref{Fig:DCR_Differences}b). This change in scattering probability has an effect on the effective penetration depth, leading to a change in $U_{max}$. Material parameters also change with $T_c$ \cite{tinkham_introduction_2015}, significantly affecting the vortex energy $\varepsilon_0$, leading to changes in $U_{max}$. We propose a general expression for the two component vortex energy ($\epsilon_0'$) combining London theory and results from TDGL
\begin{equation}
    \epsilon_0' = \frac{\Phi_0^2d}{4\pi\mu_0\lambda^2(a)}(1+\gamma\eta)
    \label{Eq:vortex energy}
\end{equation}
\begin{equation}
    a = \begin{cases}
        1.3586-0.0326T_c, & T_c\geq 11K \\
        1, & T_c<11K
    \end{cases}
    \label{Eq:interband scattering}
\end{equation}
\begin{equation}
    \lambda_{eff}^{-2}(a) = a\lambda^{-2}_\pi+(1-a)\lambda^{-2}_\sigma
\end{equation}
where $\lambda_{eff}$ is the effective magnetic penetration depth, $\eta$ is the interband Josephson coupling, and $\gamma=-1.2275$ is a fitting parameter which can be positive or negative. The expression in Eq.\ \ref{Eq:vortex energy} comes from changes to the vortex energy versus $\eta$ found using TDGL \cite{geurts_vortex_2010}. The expression for $a$ comes from a linear reduction in the $\sigma$ bandgap with decreasing $T_c$\cite{putti_observation_2006}, and taking clean MgB$_2$ to have $a=0.1$\cite{kim_reflection_2002}. Using normal state resistivities \cite{charaev_single-photon_2024,velasco_high-operating-temperature_2016,velasco_high-operating-temperature_2017} and superconducting bandgaps \cite{putti_observation_2006} from experiments along with Eq.\ \ref{Eq:DCR} and Eq.\ \ref{Eq:London}, we find the MgB$_2$ dark count rate current dependence in Fig.\ \ref{Fig:DCR_Differences}c for $T  = 4K$. Note that the dark count rate for SNSPDs based on type-1.5 MgB$_2$ reduces sharply as the current is decreased from the critical current. This behavior is vastly different from type-2 MgB$_2$, which has a significantly slower decrease in the dark count rate. We find that our model ($U_{max,s}$) matches closely with vortex barriers extracted from dark count rate experiments ($U_{max,e}$) as shown in Table \ref{tab1} (see Supplementary Materials). The remaining difference between theory and experiment may be explained by the complex relationship between interband coupling and impurity\cite{garaud_properties_2018}.

\begin{table}[bpht]
\caption{$U_{max}$ comparison at $I/I_c=0.98$}\label{tab1}%
\begin{tabular}{@{}lllllll@{}}
\botrule
  \textbf{Devices}  & \textbf{$\Delta_\sigma(0)$} & \textbf{$\Delta_\pi(0)$} & \textbf{$\rho_n$} & \textbf{$U_{max,s}$} & \textbf{$U_{max,e}$} \\
\hline
\textbf{MgB$_2$, $T_c = 37.6K$ \cite{charaev_single-photon_2024}}    & 6.2 meV  & 1.7meV & 2.5 $\mu\Omega\cdot cm$ & 114.9meV & 73.19meV   \\
\textbf{MgB$_2$, $T_c = 30.7K$ \cite{velasco_high-operating-temperature_2016}}    & 4.5 meV  & 2meV & 100 $\mu\Omega\cdot cm$ & 0.7882meV & 0.929meV   \\
\textbf{MgB$_2$, $T_c = 21.9K$ \cite{velasco_high-operating-temperature_2017}}\hspace{10pt}    & 3 meV  & 1.5meV & 120 $\mu\Omega\cdot cm$ & 1.022meV & 0.934meV   \\
\hline
\end{tabular}
\end{table}

\begin{figure}[tb!]
\centering
\includegraphics[width=\textwidth]{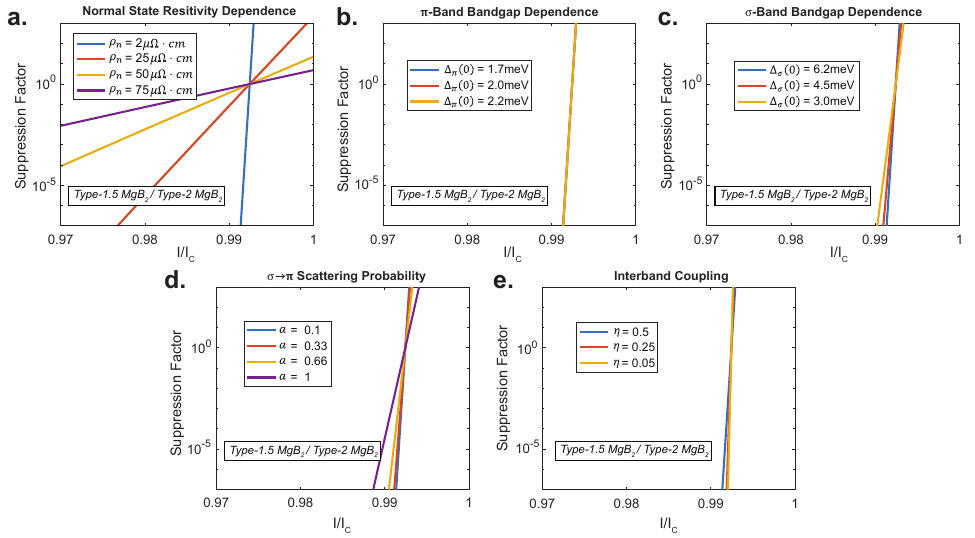}
\caption{Suppression factor dependence on material and multiband parameters. (a) Dependence of suppression factor on normal state resistivity $\rho_n$. (b) Dependence of suppression factor on bandgap of $\pi$-band $\Delta_\pi(0)$. (c) Dependence of suppression factor on bandgap of $\sigma$-band $\Delta_\sigma(0)$. (d) Dependence of suppression factor on scattering probability $a$. (e) Dependence of suppression factor on interband coupling $\eta$.}
\label{Fig:SuppressionFactorDependence}
\end{figure}

To more easily compare the dark count rate of different devices, we define a new metric from Eq. \ref{Eq:DCR} called the Suppression Factor (SF).
\begin{equation}
    SF = \frac{D_1/\alpha_1}{D_2/\alpha_2} = exp\left(\frac{U_{max,2}-U_{max,1}}{k_BT}\right)
\end{equation}
This metric measures the suppression of the dark count rate of device 1 ($D_1$) with respect to device 2 ($D_2$) normalized to the attempt rate ($\alpha_1$, $\alpha_2$). Therefore, if the devices are at the same temperature, the metric measures the reduction in dark count rate due to the difference in vortex crossing barrier $U_{max}$. In Fig.\ \ref{Fig:DCR_Differences}d, we find the dark count rate of a type-1.5 SNSPD is significantly suppressed compared to a type-2 SNSPD at high bias currents. This suppression remains even if we consider the single band case for vortex energy (i.e.\ $\eta = 0, a =1$). We also find that this suppression is significantly greater than the suppression between two type-2 SNSPDs with a similar change in $T_c$. Therefore, the change in suppression does not appear to be coming from the increase in $T/T_c$ as $T_c$ decreases. Instead, the dark count suppression appears to be caused by differences in the behavior of type-1.5 and type-2 SNSPDs. 

The suppression is strongly influenced by the vortex energy and therefore the material parameters and multiband effects. In Fig.\ \ref{Fig:SuppressionFactorDependence} we plot the dependence of the type-1.5 dark count rate suppression on normal state resistivity, $\pi$ bandgap, $\sigma$ bandgap, scattering probability, and interband coupling. The ranges for material parameters are chosen based on those typically found in experiment. We find that the change in normal state resistivity has the largest effect on suppression factor.

As we have demonstrated, the multiband nature of MgB$_2$ can lead to novel device physics. Through TDGL simulations, we have found that two-vortex clusters can nucleate from hotspots in MgB$_2$ and destroy the superconducting state. We have also found that the dark counts present in type-1.5 MgB$_2$ are significantly suppressed compared to type-2 MgB$_2$. This suppression will have a significant affect on SNSPD sensitivity at longer wavelengths or at increased operating temperatures. Experiments on type-1.5 superconductors present a clear next step in improving existing superconducting devices.

\begin{acknowledgments}
This work was funded by the DARPA SynQuaNon program. SL and PVB would like to express their gratitude to Charlsey R. Tomassetti and Elena R. Margine for providing their custom EPW code, which was utilized in calculating the band-resolved electron-phonon coupling matrix for MgB$_2$.
\end{acknowledgments}

\section*{Conflict of Interest Statement}
The authors have no conflicts to disclose.
\section*{Author Contributions}
D.H. constructed the TDGL models with inputs from Z.J. and S.B.. L.B. constructed the vortex crossing model with inputs from Z.J., S.B. and D.H.. S.L. developed the density functional theory code with inputs from P.V.B. and S.B.. Z.J. supervised the project. L.B., D.H. and Z.J. wrote the manuscript with inputs from S.B., S.L., H.X.T., and P.B.. All authors discussed the results and contributed towards writing the manuscript.
\section*{Data Availability Statement}
The data that support the findings of this study are available from the corresponding author upon reasonable request.
\bibliography{Bibliography.bib}

\end{document}


\externaldocument[supp-]{Manuscript}

\title{Supplementary Material for\\
Type-1.5 SNSPD: Interacting vortex theory of two bandgap superconducting single photon detectors}

\author[1]{Leif Bauer$^*$}
\author[1]{Daien He$^*$}
\author[1]{Sathwik Bharadwaj}
\author[2]{Shunshun Liu}
\author[2]{Prasanna V. Balachandran}
\author[1]{Zubin Jacob}

\affil[1]{The Elmore Family School of Electrical and Computer Engineering, Purdue University, West Lafayette, 47907, IN, USA}
\affil[2]{Department of Materials Science and Engineering, University of Virginia, Charlottesville, VA, 22903, USA}
\affil[*]{These authors contributed equally to the work}

\renewcommand{\theequation}{S\arabic{equation}}
 
\renewcommand{\figurename}{Fig.}
\renewcommand{\thefigure}{S\arabic{figure}}
\renewcommand\thesection{\Alph{section}}

\maketitle

\section{Density functional theory calculations}
\subsection{NbN}
Density functional theory (DFT) calculations were performed using the open-source plane-wave pseudopotential code \textsc{QUANTUM ESPRESSO} \citeSM{QE-2009,QE-2017}. The cubic crystal structure of NbN (space group, $Fm\bar{3}m$) was obtained from the ICSD database (ICSD Collection Code 982) \citeSM{christensen1977preparation, zagorac2019recent}. Prior to electronic structure calculations, the geometry was optimized using fully-relativistic SG15 optimized norm-conserving Vanderbilt (ONCV) pseudopotentials with the Perdew–Burke–Ernzerhof (PBE) exchange-correlation functional within the generalized gradient approximation (GGA) \citeSM{Perdew1996, schlipf2015optimization}. Fully relativistic pseudopotentials allow for the treatment of spin-orbit coupling (SOC) effects in the electronic structure calculations.

Electronic band structures and phonon dispersion curves were calculated based on the optimized structure. A kinetic energy cutoff of 70~Ry was used for the plane-wave basis set. The Brillouin zone was sampled using a dense $12\times12\times12$ $\Gamma$-centered Monkhorst–Pack (MP) mesh for the self-consistent field (SCF) calculation \citeSM{Monkhorst_Pack}, followed by a coarser $4\times4\times4$ MP mesh for the non-self-consistent field (NSCF) calculations required for the electron-phonon coupling calculation. Phonon calculations were performed using a $4\times4\times4$ $\mathbf{q}$-point mesh. 
The electron-phonon coupling and the superconductivity calculations are performed in the \textsc{EPW code} \citeSM{epw2016,epw2023} with a finer MP mesh of $40\times40\times40$ for the electron and phonon grid.

A relatively large cold smearing width of 0.18~Ry based on the Marzari-Vanderbilt-DeVita-Payne scheme was applied in the SCF calculations. \citeSM{marzari1999thermal} While a large smearing width is beneficial for stabilizing phonon modes and preventing imaginary phonon frequencies \citeSM{babu2019electron}, it unfortunately results in an overestimation of superconducting properties like the superconducting gap and the Bardeen–Cooper–Schrieffer (BCS) critical temperature $T_c$.

\subsection{MgB$_2$}
The hexagonal crystal structure of MgB$_2$ (space group, $P6/mmm$) was taken from Nagamatsu et al. \citeSM{Nagamatsu2001} and optimized using ultrasoft pseudopotentials \citeSM{Vanderbilt1990} within the PBE exchange-correlation functional \citeSM{Perdew1996}. All subsequent electronic structure and phonon calculations were performed using ONCV pseudopotentials \citeSM{VANSETTEN201839,oncv} within the local density approximation (LDA). A kinetic energy cutoff of 60~Ry was used for the plane-wave basis set. The Brillouin zone was sampled using an $18\times18\times16$ $\Gamma$-centered MP mesh for the SCF calculations, and a $6\times6\times6$ MP mesh for the NSCF calculations. Phonon calculations were performed on a $6\times6\times6$ $\mathbf{q}$-point mesh. A dense $80\times80\times80$ $\mathbf{k}$- and $40\times40\times40$ $\mathbf{q}$-meshes were used to compute electron–phonon coupling, and band-resolved Migdal–Eliashberg spectral function in the \textsc{EPW code}.

The band-resolved anisotropic Migdal-Eliashberg spectral function can be written as \citeSM{Margine2012}:

\begin{equation}
\alpha^2F({\bf k},{\bf k}',\omega) = N_{\rm F} \sum_{\nu} | g_{\nu{\bf k}{\bf k'}}|^2 \delta(\omega-\omega_{{\bf k}-{\bf k}',\nu}).
\label{eq:eq1}
\end{equation}

\noindent where the $ g_{\nu{\bf k}{\bf k'}}$ is the electron-phonon coupling matrix, the $N_{\rm F}$ is the density of electronic states at the Fermi level, and ${\bf q} = {\bf k} - {\bf k}'$. To resolve this spectral function into specific bands, it is necessary to distinguish between ${\bf k}$, ${\bf k}'$, and ${\bf q}$ for different electronic bands \citeSM{kafle2022}.  
Such band-resolved anisotropic Migdal-Eliashberg spectral function can then be represented by $\alpha^2F(J, J',\omega)$, which is updated from $\alpha^2F({\bf k},{\bf k}',\omega)$. In this updated notation, ${\bf k} \in J$ and ${\bf k}' \in J'$, where $J$ and $J'$ represent either the $\sigma$-band , $\pi$-band , or both. When both $J$ and $J'$ refer to the same band 
(i.e., either the $\sigma$-band or the $\pi$-band ), then $\alpha^2F(J, J',\omega)$ reflects the intra-band coupling through the diagonal elements  ($\lambda_{\sigma\sigma}$ and $\lambda_{\pi\pi}$). On the other hand, when $J$ and $J'$ refer to different bands, then the spectral function reflects the inter-band coupling through the off-diagonal elements ($\lambda_{\sigma\pi}$ and $\lambda_{\pi\sigma}$). The electron-phonon coupling matrix $\lambda_{JJ'}$ can then be written in the form of \ref{eq:eq2} \citeSM{Sanna2012}:

\begin{align}
    \lambda_{JJ'} = 2 \int \frac{\alpha^2F(J, J',\omega)}{\omega} d\omega
    \label{eq:eq2}
\end{align}

\noindent In the MgB$_2$ case, the band-resolved anisotropic electron-phonon coupling strength will consist of four elements: $\lambda_{\sigma\sigma}$, $\lambda_{\sigma\pi}$, $\lambda_{\pi\sigma}$, and $\lambda_{\pi\pi}$.

\section{Determine $U_{max}$ using TDGL}
\subsection{Single-band Time-Dependent Ginzburg-Landau Model}
Finding a saddle point state in the vortex free energy corresponds to determining the maximum barrier to vortex crossing in the superconducting nanowire with a bias current\citeSM{qiu_numerical_2008, bulaevskii_vortex-induced_2011, PhysRevLett.83.2409}. In time-dependent Ginzburg-Landau (TDGL) theory, the vortex saddle point is defined as the energy barrier that separates two local minima or metastable current-carrying states. It is equivalent to stationary solutions of a vortex at specific positions along the wire width, where its self-potential and Magnus effect cancel each other out \citeSM{vodolazov_saddle_2012}. Therefore, by finding stationary solutions of the order parameter $\psi$ in TDGL and then calculating the corresponding free energy, we can determine the vortex energy barrier $U_{max}$. 

To identify the localized saddle point state in the TDGL model, a pinning potential is applied to the vortex to select a position $x$ along the width of the nanowire \citeSM{vodolazov_saddle_2012,PriourDefect}. At the saddle point position, the vortex self-potential is compensated by the Lorentz force from the bias current $I_b$. Thus, the net force acting on the vortex becomes zero. By changing the bias current, we can find the saddle point state and thus vortex barrier for vortices at various positions of $x$. 

The normalized TDGL equations for conventional single-band superconductors (i.e. NbN, WSi, etc.) are given by
  \begin{equation}
     \left(\frac{\partial}{\partial t}+i\kappa\varphi\right)\psi=-\left(\frac{i}{\kappa}\vec{\nabla}+\vec{A}\right)^2\psi+\left(1-\frac{T}{T_c}\right)\psi-\left|\psi\right|^2\psi
     \label{eq:order_parameter_fin}
 \end{equation}
 \begin{equation}
     \sigma\left(\frac{\partial\vec{A}}{\partial t}+\vec{\nabla}\varphi\right)=\frac{1}{2i\kappa}\left(\psi^\ast\vec{\nabla}\psi-\psi\vec{\nabla}\psi^\ast\right)-\left|\psi\right|^2\vec{A}-\vec{J_b}-\vec{\nabla}\times(\vec{\nabla}\times\vec{A}-\vec{B_a})
     \label{eq:A_eqn_fin}
 \end{equation}
where $\kappa=\lambda/\xi$ denotes the ratio of penetration depth $\lambda$ to coherence length $\xi$, $\vec{A}$ is the magnetic vector potential, $\sigma$ is the conductivity of the normal current, $\vec{J_b}$ is the bias current density, and $\vec{B_a}$ is the applied magnetic field. We use the following boundary condition for current flow
\begin{equation}
\psi^*\vec{\nabla}\psi\cdot\vec{n}=i\vec{J_b}
\end{equation} 
at the boundaries with source currents, and $\vec{\nabla}\psi\cdot\vec{n}=0$ at the wire edges.

The GL free energy in its dimensionless form is given by the following equation.
\begin{equation}
     F=\int\left[\left|\psi^*(\vec{\nabla}/i-\vec{A})\psi\right|^2-|\psi|^2+\frac{\kappa^2}{2}|\psi|^4+\left|\vec{\nabla}\times\vec{A}-\vec{B_a}\right|^2\right]d^3x
     \label{eq:GLfree}
 \end{equation}
In the main text, we use the following equation for the single band free energy
\begin{equation}
    F = \int\left( F_1+F_m\right) d^3x
\end{equation}
where
\begin{equation}
    F_1 = \left|\psi^*(\vec{\nabla}/i-\vec{A})\psi\right|^2-|\psi|^2+\frac{\kappa^2}{2}|\psi|^4
\end{equation}
and
\begin{equation}
    F_m = \left|\vec{\nabla}\times\vec{A}-\vec{B_a}\right|^2
\end{equation}
which is equivalent to Eq. \ref{eq:GLfree}. The order parameter $\psi$ is calculated using TDGL equations by transforming Eq. \ref{eq:order_parameter_fin} and Eq. \ref{eq:A_eqn_fin} into coupled partial differential equations. The Ginzburg-Landau free energy $F$ is normalized by a factor $\epsilon_{0,GL}$ corresponding to the vortex energy $\epsilon_0$ in the London model.


\subsection{Two-band Time-Dependent Ginzburg-Landau Model}
Given the complex vortex behavior in MgB$_2$, we need a model capable of capturing both type-1.5 and type-2 cases. We construct the two-band time-dependent Ginzburg-Landau model (2B-TDGL) based on existing theories of multi-gap superconductors \citeSM{chan_analysis_2007,fenchenko2012phase,zhitomirsky2004ginzburg}. In the 2B-TDGL model, there are two distinctive order parameters that are weakly coupled. The dimensionless 2B-TDGL equations are given by

\begin{multline}
     \left(\frac{\partial}{\partial t}+i\phi_a\right)\psi_1=-\left(-\frac{\xi_1 i}{x_0}{\vec{\nabla}}-\frac{x_0}{\lambda_1}\vec{A}\right)^2\psi_1+\left(1-\frac{T}{T_c}\right)\psi_1\\
     -\left\vert\psi_1\right\vert^2\psi_1-\eta\psi_2-\eta_1\frac{\xi_1}{\xi_2}\frac{1}{v}\left(-\frac{\xi_2 i}{x_0}{\vec{\nabla}}-v\frac{x_0}{\lambda_2}\vec{A}\right)^2\psi_2
     \label{eq:2B_order_parameter1_2}
 \end{multline}
 \begin{multline}
     {\Gamma}\left(\frac{\partial}{\partial t}+i\phi_a\right)\psi_2=-\left(-\frac{\xi_2 i}{x_0}{\vec{\nabla}}-v\frac{x_0}{\lambda_2}\vec{A}\right)^2\psi_2+\left(1-\frac{T}{T_c}\right)\psi_2\\
     -\left\vert\psi_2\right\vert^2\psi_2-\eta\ v^2\psi_1-\eta_1\frac{\xi_2}{\xi_1}v\left(-\frac{\xi_1 i}{x_0}{\vec{\nabla}}-\frac{x_0}{\lambda_1}\vec{A}\right)^2\psi_1
     \label{eq:2B_order_parameter2_2}
 \end{multline}
 \begin{multline}
     \sigma\left(\frac{x_0^2}{\lambda_1^2}\frac{\partial\vec{A}}{\partial t}+\frac{1}{\kappa_1}{\vec{\nabla}}\phi_a\right)=\frac{1}{2i\kappa_1}\left(\psi_1^\ast{\vec{\nabla}}\psi_1-\psi_1{\vec{\nabla}}\psi_1^\ast\right)+\\
     \frac{1}{2iv\kappa_2}\left(\psi_2^\ast{\vec{\nabla}}\psi_2-\psi_2{\vec{\nabla}}\psi_2^\ast\right)-\frac{x_0^2}{\lambda_1^2}\left\vert\psi_1\right\vert^2\vec{A}-\frac{x_0^2}{\lambda_2^2}\left\vert\psi_2\right\vert^2\vec{A}-{\vec{\nabla}}\times\left({\vec{\nabla}}\times\vec{A}-\vec{B_a}\right)\\
     -\eta_1\frac{i}{2}\frac{\xi_1}{\lambda_2}(\psi_1^\ast{\vec{\nabla}}\psi_2-\psi_1{\vec{\nabla}}\psi_2^\ast+\psi_2^\ast{\vec{\nabla}}\psi_1-\psi_2{\vec{\nabla}}\psi_1^\ast)-\eta_1\frac{x_0^2}{\lambda_1\lambda_2}\vec{A}(\psi_1\psi_2^\ast+\psi_2\psi_1^\ast)
     \label{eq:2B_J_density_2}
 \end{multline}
 where $\psi_1$ and $\psi_2$ are the order parameters corresponding to the first and second band \citeSM{chan_analysis_2007}. For MgB$_2$, we have $\psi_1=\psi_{\sigma}, \psi_2=\psi_{\pi}$. $x_0$ is the geometry constant, ${\Gamma}$ is the normalized damping constant, $v=\lambda_2\xi_2/\lambda_1\xi_1$ is the product ratio of two bands' penetration depth $\lambda_i$ and coherence length $\xi_i$. The parameter $\eta$ is the coupling parameter describing the interband interactions, and $\eta_1$ is the gradient of $\eta$. The interband coupling $\eta$ is calculated based on the electron-phonon coupling strength $\lambda_{JJ'}$. The coupling before normalization is given by
 \begin{equation}
     \epsilon = \frac{\lambda_{\sigma\pi}N_\sigma}{\lambda_{\sigma\sigma}\lambda_{\pi\pi}-\lambda_{\sigma\pi}\lambda_{\sigma\pi}} 
 \end{equation}
 where $N_\sigma$ is the $\sigma$-band partial density of states. The normalized coupling is given by
 \begin{equation}
     \eta = \frac{\epsilon}{|\alpha_\sigma|}\sqrt{\frac{\beta_\sigma|\alpha_\pi |}{\beta_\pi|\alpha_\sigma|}}
 \end{equation}
 \begin{equation}
     \alpha_\sigma = N_\sigma\left[\lambda_{\pi\pi}/\eta'-\ln{(T_ce^{1/\bar{\lambda}}/T)}\right]
 \end{equation}
  \begin{equation}
     \alpha_\pi = N_\pi\left[\lambda_{\sigma\sigma}/\eta'-\ln{(T_ce^{1/\bar{\lambda}}/T)}\right]
 \end{equation}
\begin{equation}
     \beta_\sigma = \frac{7\zeta(3)N_\sigma}{16\pi^2T_c^2}
 \end{equation}
  \begin{equation}
     \beta_\pi = \frac{7\zeta(3)N_\pi}{16\pi^2T_c^2}
 \end{equation}
where $\bar{\lambda}$ is the maximum eigenvalue of the electron-phonon coupling matrix, and $\zeta(3)$ is the Riemann zeta function for $s=3$\citeSM{fenchenko2012phase,chan_analysis_2007}.

In Eq. \ref{eq:2B_order_parameter1_2}, \ref{eq:2B_order_parameter2_2}, \ref{eq:2B_J_density_2} above, we use an extension of the zero electric potential gauge choice where $\phi$ is replaced with $\phi_a$. Thus the normal current density at the boundary with source currents must satisfy
 \begin{equation}
     -\sigma\left(\frac{x_0^2}{\lambda_1^2}\frac{\partial\vec{A}}{\partial t}+\frac{1}{\kappa_1}{\vec{\nabla}}\phi_a\right)\cdot\boldsymbol{n}=\vec{J_b}\cdot\boldsymbol{n}\qquad on\quad\partial\Omega.
 \end{equation}
Since the normal component of $\frac{\partial\vec{A}}{\partial t}$ always vanishes on the boundary no matter which gauge choices we use, we can denote that there exists an electrical potential $\phi_a$ such that 
\begin{equation}
    -\frac{\sigma}{\kappa_1}{\vec{\nabla}}\phi_a=\vec{J_b}\qquad on\quad\partial\Omega.
    \label{2B_Ja}
\end{equation}
Other boundary conditions on $\partial\Omega$ are nearly the same as the one-band TDGL as shown below:
 \[\left(-\frac{\xi_1 i}{x_0}{\vec{\nabla}}\psi_1-i\eta_1\frac{1}{v}\frac{\xi_2}{x_0}\vec{\nabla}\psi_2\right)\cdot\boldsymbol{n}=0\]
 \[\left(-\frac{\xi_2 i}{x_0}{\vec{\nabla}}\psi_2-i\eta_1v\frac{\xi_1}{x_0}\vec{\nabla}\psi_1\right)\cdot\boldsymbol{n}=0\]
 \[\vec{\nabla}\times\vec{A}=\vec{B_a}\]
 \[\vec{A}\cdot\boldsymbol{n}=0\]

To determine the saddle point state in two-band superconductors (e.g. MgB$_2$), we follow a similar approach as what we did for single-band superconductors. We perform both the single-vortex crossing case and the two-vortex crossing in each band of MgB$_2$. Once we reach the saddle point states ($\delta\psi/\delta t = 0$) under various bias currents, the GL free energy is calculated based on two order parameters $\psi_\sigma, \psi_\pi$ determined from the two-band TDGL model. In two-band TDGL, the free energy is given by
\begin{equation}
     F(\psi_\sigma,\psi_\pi)=\int(F_\sigma+F_\pi+F_{\sigma\pi}+F_m) d^3x
     \label{eq:F_twoband}
 \end{equation}
where
\begin{equation}
F_\mu=\left|\psi_\mu^*(\vec{\nabla}/i-\vec{A})\psi_\mu\right|^2-|\psi_\mu|^2+\frac{\kappa_\mu^2}{2}|\psi_\mu|^4\quad\mu=\sigma,\pi
\label{eq:F_twoband2}
\end{equation}
 \[F_{\sigma\pi}=\eta(\psi_\sigma^*\psi_\pi+\psi_\pi^*\psi_\sigma)\]
 \[F_m=|\vec{\nabla}\times\vec{A}-\vec{B_a}|^2\]

\section{Vortex Energy}
The London vortex energy $\epsilon_0$ can be derived from London theory to be
\begin{equation}
    \epsilon_0 = \frac{\Phi_0^2d}{4\pi\mu_0\lambda^2}
\end{equation}
where $\Phi_0=h/2e$ is the magnetic flux quantum, $d$ is the film thickness, $\mu_0$ is the vacuum permeability, and $\lambda$ is the  magnetic penetration depth. The London vortex energy is proportional to the magnetic and kinetic energy of a vortex in an infinite thin film \citeSM{kogan_interaction_2007}. On the other hand, the vortex energy in Ginzburg-Landau theory is calculated from the LAMH phase slip energy using Ginzburg-Landau relations \citeSM{tinkham2002quantum}. 
\begin{equation}
     \epsilon_{0,GL} = \frac{\sqrt{6}\hbar}{2e}\frac{\Phi_0}{2\sqrt{2}\pi\xi}\frac{(2/3)^{3/2}wd}{\mu_0\lambda^2} = \epsilon_0\frac{2^{3/2}W}{3\pi\xi}
     \label{eq:GL_main}
 \end{equation}
The factors $\epsilon_0$ and $\epsilon_{0,GL}$ are used to normalize and compare the current-dependent vortex crossing barriers from London and Ginzburg-Landau models.

\section{MgB$_2$ transition from type-1.5 to type-2}
The behavior of normal domain formation in superconductors is primarily determiend by the Ginzburg-Landau criterion, i.e. 
\begin{equation}
    \begin{cases}
        \frac{\lambda}{\xi} \lesssim \frac{1}{\sqrt{2}},  & Type-1 \\
        \frac{\lambda}{\xi} \gtrsim \frac{1}{\sqrt{2}} , & Type-2
    \end{cases}
\end{equation}
where $\lambda$ is the magnetic penetration depth and $\xi$ is the coherence length. As we've previously mentioned, clean MgB$_2$ exhibits type-1.5 behavior due to the type-1 $\pi$-band and type-2 $\sigma$-band. However, as impurities are added and the critical temperature decreases, MgB$_2$ transitions into a type-2 superconductor. This occurs due to the type-1 $\pi$-band transitioning to type-2. Theoretical calculations of the $\pi$-band coherence length tell us that the clean $\pi$-band Ginzburg-Landau parameter is close to the transition between type-1 and type-2 at $0.92/\sqrt{2}$ \citeSM{tinkham_introduction_2015,moshchalkov_type-15_2009}. Therefore even a small increase in $\lambda_\pi$ will lead to a transition from type-1.5 to type-2.
From BCS theory we find that the magnetic penetration depends on the normal state resistivity $\rho_n$ and superconducting bandgap at $T=0$, $\Delta_i(0)$. In multiband superconductors the magnetic penetration depth is given by
\begin{equation}
    \lambda_i^2(0) = \frac{\hbar\rho_n}{\pi\mu_0\Delta_i(0)}
\end{equation}
\begin{equation}
    \lambda^{-2}=a\lambda_2^{-2}+(1-a)\lambda_1^{-2}
\end{equation}
where $\lambda_i$ is the magnetic penetration depth of band $i$, and $a$ is the interband scattering probability. Since the normal state resistivity increases as $T_c$ decreases \citeSM{putti_intraband_2007}, MgB$_2$ should quickly pass into the type-2 regime as $T_c$ is reduced. However, since a reduction in $T_c$ leads to a transition from the clean to dirty limit \citeSM{tinkham_introduction_2015,tarantini_effects_2006}, it is difficult to mark the exact point at which the transition occurs. In the dirty limit, the Ginzburg-Landau parameter is given by
\begin{equation}
    \kappa = 0.715\frac{\lambda}{l}
\end{equation}
where $l$ is the mean free path. Assuming that MgB$_2$ enters the dirty limit as $T_c$ is reduced, we can expect at $35<T_c<37$ the decreasing mean free path \citeSM{tarantini_effects_2006} and increasing $\lambda$ will lead to the transition. Therefore, due to the sharp increase in magnetic penetration depth, we expect by $T_c\lesssim35$ the MgB$_2$ SNSPD will be in the type-2 regime.

\section{Extraction of Vortex Barrier from Dark Count Rate Experiments}
The vortex barrier can be extracted from dark count rate experiments by fitting an exponential to the dark count rate. Taking the natural log of Eq. 2 we have
\begin{equation}
    ln(D) = c_1-c_2*\max_{x_\nu}\left[\ln\left(\frac{2W}{\pi\xi}\right)\sin(\pi x_\nu)-\frac{I}{I_c}\frac{2Wx_\nu}{\exp(1)*\xi}\right]/k_BT
\end{equation}
where $c_1,c_2$ are fitting parameters with $\alpha=e^{c_1}$. Then, the second term times $k_BT$ is $U_{max}$. Therefore, $U_{max}$ can be extracted given the experiment parameters $\xi,W,I/I_c$, and $T$.
\section{Simulation Flowchart}
Below we demonstrate how DFT and material parameters are combined to obtain TDGL results and ultimately dark count rate and suppression factor predictions.
\begin{figure}[ht!]
\centering
\includegraphics[width=\textwidth]{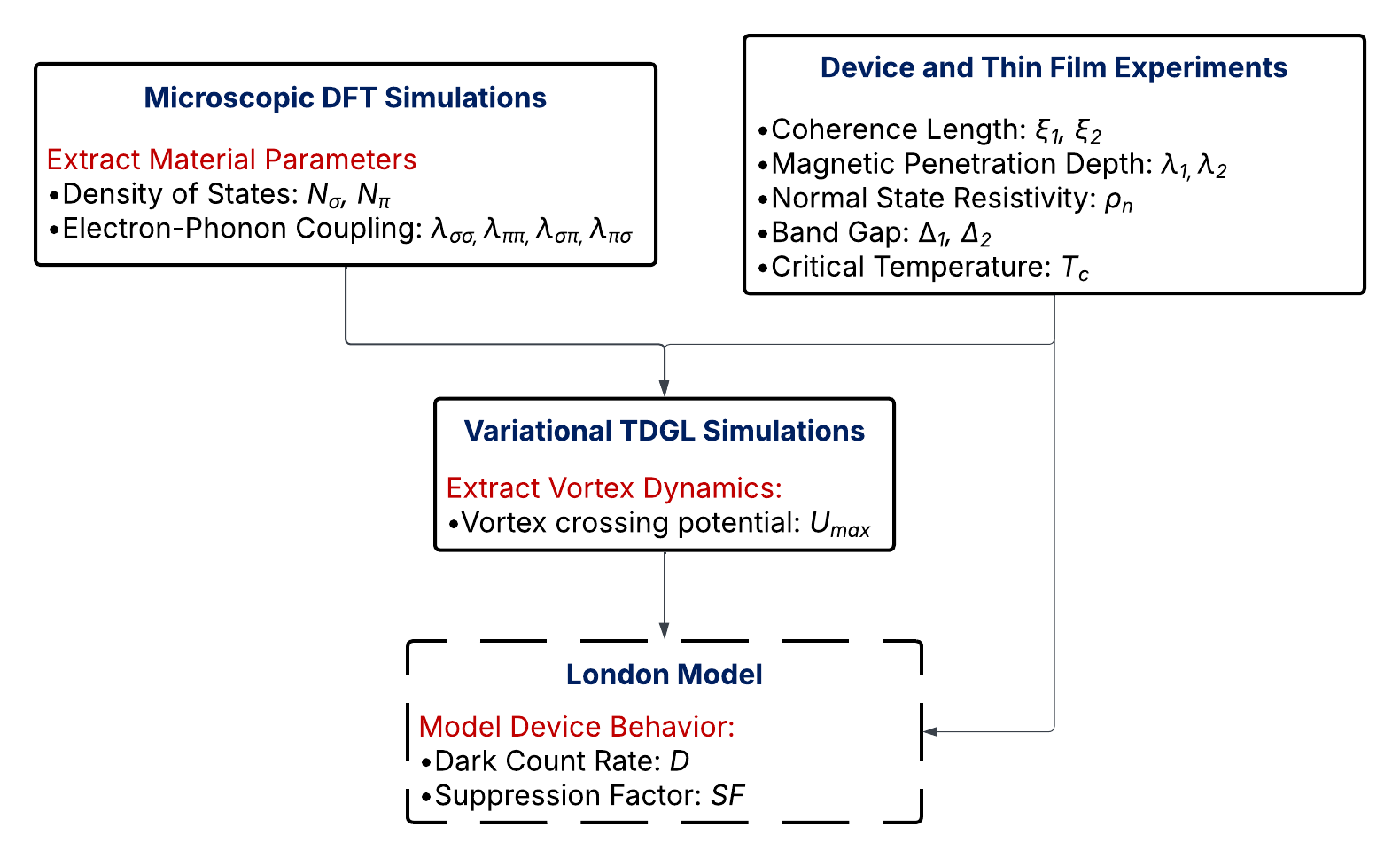}
\caption{Flowchart of simulations. Electron-phonon coupling matrix and density of states are obtained from DFT. Other material parameters come from experiment and are used in TDGL and London theory calculations of $U_{max}$, dark count rate, and suppression factor.}
\label{Fig:flowchart}
\end{figure}

\bibliographystyleSM{unsrt}
\bibliographySM{Bibliography}